\documentclass[eqsecnum]{revtex4} 

\usepackage{psfig}
\begin{document}
\title{Thermodynamic instabilities in one dimension:
correlations, scaling and solitons}
\author{Thierry Dauxois$^{1}$\thanks{Thierry.Dauxois@ens-lyon.fr},
Nikos Theodorakopoulos$^{2}$\thanks{nth@eie.gr} and
Michel Peyrard$^{1}$\thanks{Michel.Peyrard@ens-lyon.fr} }
\affiliation{$^{1}$ Laboratoire de Physique, UMR-CNRS 5672,
ENS Lyon, 46 All\'{e}e d'Italie, 69007 Lyon, France \\
$^{2}$ Theoretical and Physical Chemistry Institute,
National Hellenic Research Foundation,\\
Vasileos Constantinou 48, 116 35 Athens, Greece}
\date{\today}
\begin{abstract}

Many thermodynamic instabilities in one dimension (e.g. DNA
thermal denaturation, wetting of interfaces) can be described in
terms of simple models involving harmonic coupling between nearest
neighbors and an asymmetric on-site potential with a repulsive
core, a stable minimum and a flat top. The paper deals with the
case of the Morse on-site potential, which can be treated exactly
in the continuum limit. Analytical expressions for correlation
functions are derived; they are shown to obey scaling; numerical
transfer-integral values obtained for a discrete version of the
model exhibit the same critical behavior. Furthermore, it is shown
in detail that the onset of the transition can be characterized by
an entropic stabilization of an -otherwise unstable-,  nonlinear
field configuration, a soliton-like domain wall (DW) with
macroscopic energy content. The statistical mechanics of the DW
provides an exact estimate of the critical temperature for a wide
range of the discretization parameter; this suggests that the
transition can be accurately viewed as being "driven" by a
nonlinear entity.

\end{abstract}
\maketitle

\author{N. Theodorakopoulos, T. Dauxois, M. Peyrard}
\date{\today}
\maketitle

\section{INTRODUCTION}
\label{introduction}

In a seminal paper published a quarter-century ago, Krumhansl
and Schrieffer \cite{KRUMSCHRIEF}
explored the possibility that nonlinear excitations (solitons)
might drive structural phase transitions.
They succeeded in
identifying signatures of "domain walls (DW)" in the thermodynamic
properties of the one-dimensional
$\phi^{4}$ model and gave a phenomenological
account of a salient feature of structural phase transitions, the
appearance of a central peak in the dynamical spectrum.
However, important issues remained open. In particular, because the
demonstration was limited to a system which does not
exhibit a genuine (i.e. finite-temperature) thermodynamic phase transition,
there was no possibility to explore the role of exact, or nearly
exact, nonlinear field configurations in determining
critical behavior, static or dynamic.

A class of one-dimensional models, which has been proposed
to describe a wide variety of thermodynamic instabilities, such
as thermal DNA denaturation
\cite{PB,dauxpeyr1,dauxpeyr2},
wetting of interfaces \cite{KroLip}, and
other similar phenomena, could prove to be of interest in the
above context.
Typically, such models include a short-range
interaction between near neighbors,
and an on-site potential with a single minimum and a flat-top;
in the context of DNA denaturation, as the temperature
increases, particles begin to access the region of the flat
top at an increasing rate, until, at a finite temperature, a macroscopic
instability occurs, with a divergence in the average displacement
(the "effective length" of the hydrogen bond which holds the
double-stranded chain together), a divergent correlation length, and,
generally, all the characteristics of a thermodynamic phase transition.
In the simplest case, that of a harmonic interaction and a Morse
on-site potential, the thermodynamics can be calculated exactly
by using functional integral techniques \cite{KRUMSCHRIEF}
and mapping the problem to the quantum mechanics of the
Morse oscillator. The thermodynamic transition corresponds to
the quantum mechanics of the disappearance of the last bound
state.

In this paper, we will show that soliton physics can offer a
correct interpretation of such thermodynamic instabilities. In
particular, a relatively brief calculation can show us {\em why}
and {\em when} a thermodynamic transition occurs. The logic can be
summarized as follows: An exact, static nonlinear configuration of
the continuum field can be found, which "interpolates" between the
low- and the high-temperature phase. We interpret this to a be a
DW.  The total energy of the configuration diverges, because each
site which is in the high-temperature phase contributes a finite
amount of bond (and elastic) energy. The infinity is in a sense a
"minimal requirement": as long as the DW energy is finite, no
phase transition can occur in one dimension \cite{Landau}. A
closer look at the properties of the DW is more revealing. In
contrast to other Klein-Gordon field theories with a finite-energy
kink, there is no zero-frequency Goldstone mode. It costs energy
to shift the position of DW so that it can include more sites at
the high-temperature phase. Conversely, there is an energy gain by
"zipping back" to the low-temperature phase. At finite
temperatures, the picture must be corrected to include entropic
effects. The "phonon cloud" which accompanies the DW provides an
entropic gain which ultimately balances the energy cost of
extending the DW to macroscopic scale. The formation of a
thermally stable DW is thus seen to "drive" the instability. At
higher temperatures, it is entropically advantageous for a DW to
extend itself by "unzipping" more sites towards the high
temperature phase. The picture sketched above, which concentrates
entirely on the thermal stability of the nonlinear interface, is
more than a qualitative nonlinear scenario. It provides an
estimate of the transition temperature which (i) turns out to be
exact within the continuum field framework  and, (ii) can be {\em
nonperturbatively} extended to describe lattice systems up to
quite high levels of discretization.

The paper is structured as follows: Section II provides a full
account of exact results for the model. This is important
to establish notation and context for the discussion. Moreover,
although some results have appeared in the literature before,
they are scattered in various sources; in addition, results
for the correlation function and the susceptibility are new and
the calculation has not been reported elsewhere \cite{newprl}.
Section III deals with the soliton physics of the transition.
A brief discussion of results and further perspectives, with
special emphasis to effects due to lattice discreteness,
is given in Section IV.

\section{Exact thermodynamics}
\label{CriticalPhenomena}

\subsection{Model and notation}
The Hamiltonian of the model is \cite{dauxpeyr1}
\begin{equation}
H=\sum_{n}^{}\Biggl[ {p_n^2 \over 2 m}
+{K\over 2}(y_n-y_{n-1})^{2} + V(y_n)  + Dha y_n   \Biggr]
\label{eq:eHamiltonian}
\end{equation}
where $m$ is the reduced mass of a base pair, $y_n$ denotes the
stretching of the hydrogen bonds connecting the two bases of the
$n^{\text{th}}$ pair and $p_n = m (dy_n/dt)$ corresponds to the
conjugate momentum of $y_n$.

\begin{figure}
\null\hskip 1truecm\psfig{figure=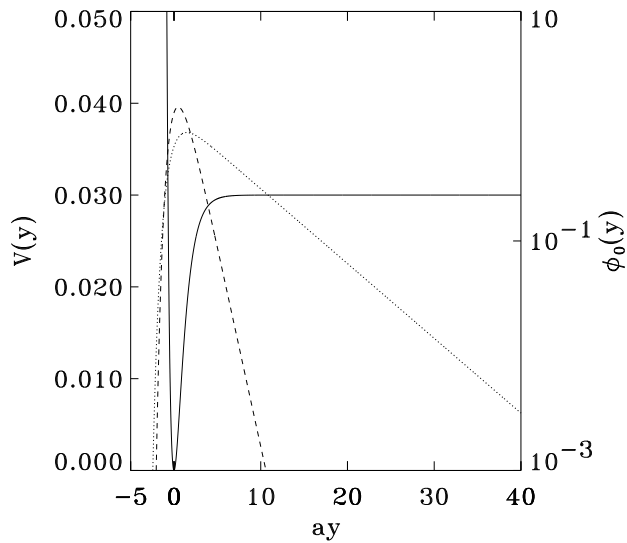,height=7truecm,width=7truecm}\hfill
\caption{The Morse potential $V(y)$ is shown (solid line) as a
function of the dimensionless field $ay$. The groundstate
wavefunction $\phi_0(y)$ (Eq.~(\ref{phizerotypeI})) is shown for
different values of the temperature: dashed line for $T=T_c/2.6$,
dotted line for $T=T_c/1.3$. Note the logarithmic scale for
$\phi_0(y)$. } \label{diffphizero}
\end{figure}

 In addition to the kinetic and nearest-neighbor potential energy terms,
(\ref{eq:eHamiltonian}) contains: (i) an on-site
potential which describes the interaction of the two bases in a pair;
the Morse potential
\begin{equation}
V(y_n) = D\ (e^{-ay_n}-1)^2\quad,
\label{eq:onsite}
\end{equation}
plotted in Fig.~\ref{diffphizero}, has been chosen because it has
the correct qualitative shape; and (ii), a field-dependent term,
which describes the effect of
a transverse, external stress $h$.

A set of parameters which has been used in the DNA denaturation
context~\cite{dauxpeyr1,dauxpeyr2} is: a dissociation energy
$D=0.03$ eV, a coupling constant $K=0.06$ eV/$\AA^2$, a spatial
scale factor of the Morse potential $a=4.5\ \AA^{-1}$, and a mass
$m=300$ a.m.u; the time scale is determined by $\omega
_{0}=(K/m)^{1/2}$.  The dimensionless ratio $R=Da^2/K$ is
traditionally used to distinguish between the "order-disorder"
(discrete, $R\gg1$) and "displacive" (continuum, $R\ll 1$) regime;
the given set of parameters corresponds to $R=10.1$.

The classical thermodynamic
properties of (\ref{eq:eHamiltonian}) can be described exactly
in terms of the transfer integral (TI) equation
~\cite {Kac,KRUMSCHRIEF}
\begin{eqnarray} \int_{-\infty}^{+\infty} d y \ e^{ - \beta \left[V(x)+V(y)+2W(x,y) +
{Dah} (y+x )\right] /2 } \
\phi_n(y)= e^{- \beta \varepsilon_n   }\
\phi_n(x) \quad,
\label{eq:transferoperator}
\end{eqnarray}
where $\beta =1/(k_{B}T)$,  $k_{B}$ is the Boltzmann
constant and $T$ is the temperature.

In general, we are interested in situations for which
$h=0$; however, the external field is useful in
practical calculations  as a
mathematical device. For example, by letting $h\to 0^{+}$
it is possible to extract the scaling behavior near
the transition; moreover, since the partition function is now
divergence-free at any $h>0$ (and numerical calculations
can in principle be performed at finite, decreasing $h$),
previous criticism of the model on
formal mathematical grounds  \cite{chinese} is addressed.

In the thermodynamic limit, the free energy per site (except for a
nonsingular term due to the integral over momenta) is
determined by the smallest eigenvalue
$\varepsilon_{0} $ of (\ref{eq:transferoperator}), i.e.,

\begin{equation}
f =  \varepsilon_{0}  - \frac{1}{2\beta}
\ln\left[ \frac{2\pi m}{\beta} \right]
\label{freeenergy1}
\end{equation}

Other thermodynamic properties of interest
are (i) the order parameter
\begin{equation}
\sigma= <y>=\int_{-\infty}^{+\infty} dy\ y\ |\phi_{0}| ^{2}\\
\label{OrderP}
\end{equation}
where $\phi_{0}$ is the normalized eigenstate corresponding to the
eigenvalue $\varepsilon_{0} $
 and (ii) the correlation function
\begin{equation}
C(j) \equiv <(y_{j}-\sigma)(y_{0}-\sigma)> \\
=    \sum_{n=1}^{+\infty} |M_{n}|^{2} e ^ { - \Delta_{n} |j|},
\label{Corrdef}
\end{equation}
where the sum runs over states other than
$\phi_{0}$,
\begin{equation}
\Delta_{n}  \equiv \beta (\varepsilon_{n} - \varepsilon_{0})
\quad ,
\label{deltandef}
\end{equation}
and the off-diagonal matrix elements $M_n$ are given (in
Dirac notation) by
\begin{equation}
M_{n} = <n|(y-<y>)|0> = <n|y|0>  \quad.
\label{mateldef}
\end{equation}

\subsection{The corresponding Morse oscillator problem}
\label{boundscatteringstates}
In the gradient expansion (continuum) approximation,
\begin{equation}\phi_n(y)=\phi_n(x)+\phi_n'(x)(y-x)+
\phi_n''(x)(y-x)^{2}/2+{\cal O}\left((y-x)^{3}\right)\quad,
\end{equation}
which is strictly valid in the temperature window $D\ll k_{B}T \ll
D/R$ \cite{GuyerMiller}, the integral equation
(\ref{eq:transferoperator}) can be well approximated by the
second-order differential equation
\begin{equation}
  \left[ - \frac{1}{2\beta^{2}K}\frac{d^{2}}{d y^{2}} +
    D(e^{-ay}-1)^{2} + Dhay \right] \phi_{n} (y) = {e_{n}}\
  \phi_{n} (y) \quad .
\label{eqmorse}
\end{equation}
where $ e_{n}= \varepsilon_{n} + \ln({2\pi }/ {\beta
K})/{2\beta}$. The one-dimensional statistical mechanics TI
problem is thus mapped~\cite{PB} to the zero-dimensional quantum
mechanics problem of the Morse oscillator.

In the following, we will list and/or derive some general
properties of (\ref{equadiff}) at $h=0$. The change of variables
\begin{equation}
z=2\delta \exp(-ay)\quad {\rm with}\quad \delta = \frac{\beta}{a} \sqrt{2DK}\quad ,
\label{defdelta}
\end{equation}
the transformation $ \phi_n(y) = e^{-z/2}\ z^{s}\ w_n(z) $ and $s=
\delta\sqrt{1- e_{n}/D} $, leads to
\begin{equation}
z {d^2w_n\over dz^2}  +  (2s+1-z) {dw_n\over dz} + n w_n = 0  \quad ,
\label{equadiff}
\end{equation}
where $
n = \delta - s -1/2$.  If $n$ is a
positive integer, the solution of (\ref{equadiff})  is a
Laguerre polynomial~\cite{morsestueckelberg}.
\subsection{Bound state(s) and associated distinct length scales}
\label{boundstates}
Noting that $\phi_n(y)$ remains
finite over the interval $[0,+\infty[$ only for positive values of $s$,
we obtain the spectrum of bound states,
\begin{eqnarray}
\label{spectrum}
\frac{e_{n}}{D} = 1 - \left( 1 - \frac{n+1/2}{\delta}
\right)^{2} \quad {\rm with}\quad
n = 0,1,\dots, E(\delta -1/2) \; ,
\end{eqnarray}
where $E(\delta -1/2)$ is the integer part of $\delta -1/2$.
It follows that as long as $\delta$ exceeds
a critical value $\delta_c=1/2$, the ground state remains
bounded. In the quantum problem, this provides
a criterion for the critical mass of a particle
below which it is driven out of the potential well by quantum
fluctuations. We note that this is a general property of
asymmetric potential wells; symmetric wells support
a bound state for any value of
$\delta$~\cite{buell,MF}.

Using~(\ref{defdelta}), this $\delta_c$ defines a critical temperature
\begin{equation}
T_{c}= \frac{2\sqrt{2KD}}{ak_{B}}
\label{Tcexact}
\end{equation}
in terms of which $\delta = T_{c}/2T$.  As the temperature
approaches $T_{c}$, the last bound state becomes less and less
localized, and $<y>$  increases sharply, indicating the "melting" of
the system.

Using (\ref{freeenergy1}) results in a free energy per site
$ f= e_{0} + f_{0}$, where
\begin{equation}
f_{0} = -\frac{1}{\beta}\ln\left( \frac{ 2\pi }{\beta \omega_{0}
}\right) \label{f0}
\end{equation}
and
\begin{eqnarray}
e_{0}  &=& D\left [1 -  \left| t \right|^{2} \right]
 \qquad \mbox{if $T<T_c$}\\
\nonumber
&=&  D \quad \mbox{otherwise}
\label{freeenergy}
\end{eqnarray}
where $t=T/T_{c}-1$  is the reduced temperature; it
follows that the entropy per site can be written as
the sum
of a non-singular part
\begin{equation}
S_{non-sing}=S(T_c)+k_B\log\left({T\over T_c}\right)
  \label{entropynonsing}
\end{equation}
and a singular part,
\begin{eqnarray}
 \label{entropysing}
S_{sing}&=&
{2D\over T_c}\ t  \qquad \mbox {if}  \qquad T<T_c \\
\nonumber &=&{0}\qquad \qquad\mbox{otherwise.}
\end{eqnarray}
We note that (\ref{entropysing}) implies a jump discontinuity of
the specific heat at $T_{c}$; this is consistent with a specific
heat critical exponent $\alpha=0$.

For  $T_{c}/3 <T<T_{c} $, the (normalized) ground state
\begin{equation}
\phi_{0} = \sqrt{ \frac{a}{\Gamma(2\delta-1) }}\ e^{-z/2}\ z^{\delta-1/2} \>,
\label{phizerotypeI}
\end{equation}
where $\Gamma$ is the gamma function, is the only bound state. The
asymptotic behavior $ \phi_{0}(y) \propto e ^{- a( \delta-1/2) y}
$ defines the spatial extent of the ground state
\begin{equation}
\lambda  = \frac {1} { a( \delta-1/2) }\quad.
\end{equation}

\begin{figure}
\null\hskip 1truecm\psfig{figure=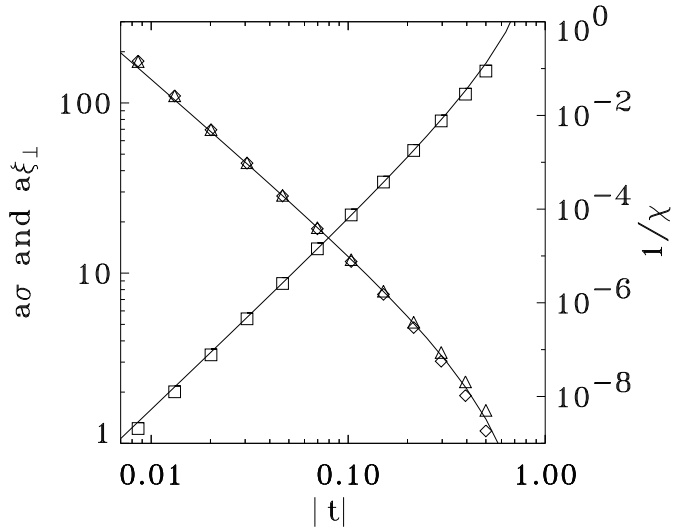,height=7truecm,width=7truecm}
\vskip  1truecm
\caption{Left scale: dependence of the order parameter $\sigma$ (diamonds)
and its fluctuations $\xi_{\perp}$ (triangles) as
a function of the reduced temperature $t$; the points
were obtained numerically using the TI method;
the solid line shows the analytical estimate, Eq.~(\ref{lamndasur2}),
with $\delta_c=0.36$.
Right scale: the susceptibility $\chi$ as a function
$t$ (squares: numerical TI; solid line: Eq.~(\ref{eqchi}) with
$\delta_c=0.36$).}
\label{sigmaandycarreI}
\end{figure}

The order parameter, obtained from Eqs. (\ref{OrderP}) and
(\ref{phizerotypeI}), is~\cite{NietoSimmons}
\begin{equation}
 \sigma = \frac{1}{a} \left[ \ln(2\delta)- \psi(2\delta-1)\right]
\end{equation}
where  $\psi$ is the digamma function;
in the vicinity of the critical temperature, this reduces to
\begin{equation}
\sigma\sim \frac{1}{a(2\delta-1)}={\lambda\over 2}= {1\over
2a\delta_c}\ {T\over T_c} \left| t \right| ^{-1}  \quad ;
\label{lamndasur2}
\end{equation}
in the language of phase transitions,
$\sigma \propto |t|^{\beta}$,   where $\beta=-1$ is the
critical exponent  for the order parameter, i.e. the order parameter
{\em diverges} at the instability. Fig.~\ref{sigmaandycarreI}
shows that the agreement of numerical TI values with
(\ref{lamndasur2})
is excellent, provided
one takes into account the numerical TI value of $\delta_c=0.36$
(corresponding to $T_{ c}=427 K$)
for the discrete system.

Local fluctuations of the order parameter are described by
\cite{NietoSimmons}
\begin{eqnarray}
  \xi_{\perp}=\sqrt{<0\ |\left(y-\sigma\right)^{2}|\ 0>}
  =\frac{\sqrt{\psi{'}\left(2\delta -1\right)}}{a} \quad \; .
\label{perp}
\end{eqnarray}

In the vicinity of the critical temperature, this reduces to
\begin{eqnarray}
\xi_{\perp}\sim \frac{1}{a (2\delta -1) }= \frac{\lambda }{2}
\label{fluct}
\end{eqnarray}
which implies  $\xi_{\perp} \propto |t|^{-\nu_{\perp}}$ with
$\nu_{\perp}=-\beta=1$. Close to the transition, the order
parameter and its fluctuations are therefore comparable, as shown
by Fig.~\ref{sigmaandycarreI}; in the wetting
literature~\cite{KroLip} they are both interchangeably referred to
as ``transverse correlation lengths''.

\begin{figure}
\null\hskip 1truecm\psfig{figure=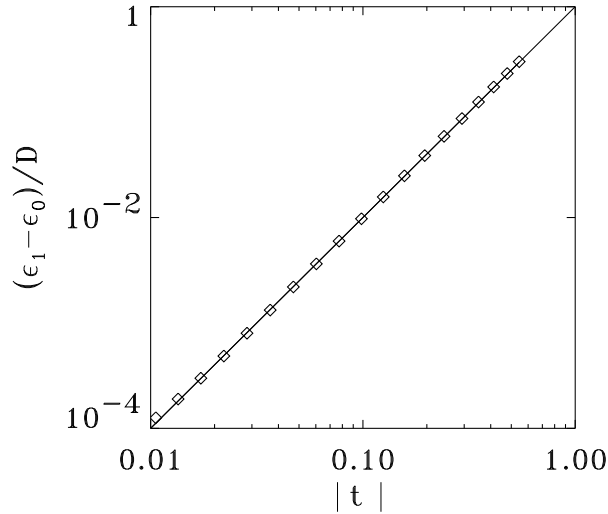,height=7truecm,width=7truecm}
\vskip 1truecm \caption{ Dependence of $(e_{1}- e_{0})/D$ on the
reduced temperature. The solid line corresponds to the exact
(continuum) result $|t|^{2}$, cf. (\ref{Eqxi}). The diamonds are
numerical TI results.} \label{e1moinseotypeI}
\end{figure}

If we denote the lattice constant by $\ell$ and identify the next-to-lowest
eigenstate with the bottom of the continuum band, we obtain the
inverse longitudinal correlation length
\begin{equation}
\frac{\ell}{\xi}_{||} = \beta(e_{1}- e_{0}) \simeq \beta(D- e_{0})
= \beta D\ |t|^2 =\frac{\beta D} { (\delta a \lambda)^{2} }
\label{Eqxi}
\end{equation}
i.e.  the corresponding critical exponent $\nu_{||}=2$. The solid
line plotted in Fig.~(\ref{e1moinseotypeI}) shows a perfect
agreement for the difference of the two first eigenvalues, between
Eq.~(\ref{Eqxi}) and numerical TI results.

\subsection{Static structure factor and susceptibility}
\label{harmoniccasetrois}

In terms of the correlation functions derived above, the static
structure factor reads
\begin{eqnarray}
S(q,T) &=& \sum_{j=-\infty}^{+\infty} e^{iqj\ell}\ C(j) \\
&=&\sum_{j=-\infty}^{+\infty}\sum_{\kappa=1}^{+\infty} |M_{\kappa}|^{2}
 e ^{iqj\ell - \Delta_{\kappa}|j|}  \\
&=&\sum_{\kappa=1}^{+\infty} |M_{\kappa}|^{2}  \frac { \sinh{\Delta_{\kappa} }     }
{ \cosh{\Delta_{\kappa}  } - \cos(q\ell)  } \; ,
\label{sqfinal}
\end{eqnarray}
where the sum extends over all the scattering states of
Eq.~(\ref{eqmorse}).
They are given by the second solution \cite{Abr_S}
of the confluent hypergeometric equation~(\ref{equadiff}) for imaginary
$s=i\kappa$. In the notation of  Eq.~(13.1.1) of Ref. \cite{Abr_S},
the solution
$w(z) = U(1/2-\delta + i\kappa, 1+2i\kappa,z)$ has the correct
asymptotic behavior: it grows as $z^{\delta-1/2-i\kappa}$ as $z \to
+\infty$ $ ({\rm i.e.,\ }y \to -\infty)$, so that the exponential
prefactor in the definition of $\phi_\kappa$ dominates the vanishing
of the wavefunction at the Morse hard core. These scattering states
correspond to plane wave superposition
\begin{equation}
w(z) \propto 1 -  e^{2i\theta(\kappa,\delta) } z ^{-2i\kappa}
\end{equation}
as $z \to 0$ $ ({\rm i.e.,\ }y \to +\infty)$ where $\theta(\kappa)
= \arg \Gamma(1+2i\kappa)  + \arg\Gamma(1/2-\delta-i\kappa).$ The
unnormalized continuum eigenstates are therefore given in the
asymptotic limit $y \to \infty$ by
\begin{equation}
\phi_{\kappa}(y) = \sin( \kappa a y + \theta(\kappa))
\label{asycont}
\end{equation}
and the corresponding eigenvalues by
\begin{equation}
e_{\kappa} = D\left(1 + {\kappa^{2}}/{\delta^{2}}\right)
\quad .
\label{scatt_ev}
\end{equation}

As $\varepsilon_{\kappa}$ and $e_{\kappa}$ differ by a quantity
that does not depend on $\kappa$, from (\ref{scatt_ev})  and
(\ref{deltandef}) we identify
\begin{equation}
\Delta_{\kappa} = \frac{\beta D}{\delta^{2}} \left[
(\delta-1/2)^{2}+ \kappa^{2}\right]\quad . \label{deltan}
\end{equation}

\smallskip

Near the critical instability and in the long-wavelength limit
$q\ell\ll1$, the static structure factor can be calculated as
follows for a given value of $q\xi_{||}$.  Due to the rapid
oscillations of the scattering eigenstates~(\ref{asycont}), matrix
elements will vanish for large values of $\kappa$ whereas, for
small $\kappa$, the phase shift can be approximated by
$\theta=-\arctan(\kappa \lambda a)$.  Since the ground state
extends to large values of $y$, for most of the interval of
integration we can use the asymptotic form (\ref{asycont}) of the
continuum states.  Normalizing  in a box of size (${\cal
O}(L^0),L/2)$ to account for the soft core, the ground and excited
states are respectively
\begin{eqnarray}
\phi_{0}(y) &\simeq& \sqrt{\frac{2}{{\lambda}}}    e^{-y/\lambda} \\
\phi_{\kappa}(y)& \simeq &\frac{2i}{\sqrt{L}} \sin( \kappa a y+\theta( \kappa))
\label{phik},
\end {eqnarray}
where $L$ is the system size. We obtain
\begin {eqnarray}
  M_{\kappa} &\simeq& i \sqrt{\frac{8\lambda^{3}}{L}}\
  \frac{\sin(\sigma+\theta)} {1 + \kappa^{2}\lambda^{2} a^{2}}
\end {eqnarray}
where $ \sigma \equiv 2 \arctan (\kappa \lambda a)=-2\theta $. This
explicitly gives $\tan(\theta+\sigma)=\kappa \lambda a$ and hence
\begin{equation}
  |M_{\kappa}|^{2} \simeq \frac{8\lambda^{3} }{L} \frac{
    \kappa^{2}\lambda^{2} a^{2} } {(1+ \kappa^{2}\lambda^{2} a^{2}
    )^{3} }
\; .
\end{equation}
In order to substitute the sum over ${\kappa}$ by an integral in Eq.~(\ref{sqfinal}),
we need the correct density of space in
$\kappa$ space; this can be obtained by setting the eigenfunction of
(\ref{phik}) equal to zero at the boundary
$x=L/2$ and leads to a density of
states $aL/2\pi$. Eq. (\ref{sqfinal}) can
be rewritten as

\begin{figure}
\null\hskip 1truecm\psfig{figure=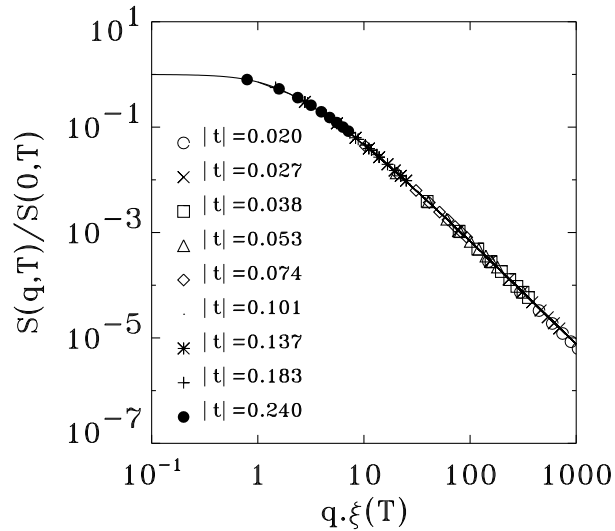,height=7truecm,width=7truecm}
\vskip 1truecm \caption{The reduced static structure factor
$S(q,T)/S(0,T)$ as a function of $q\xi$ for different values of
the reduced temperature.The solid line is the theoretical
curve~(\ref{redsqt}) whereas the symbols were obtained using the
TI method for different values of the temperature.}
\label{sqttypeI}
\end{figure}

\begin{equation}
S(q,T)= \lambda^{2} \frac{\xi_{||}}{\ell} F(q\xi_{||}) \quad,
\label{defsqtypeI}
\end{equation}
where
\begin{eqnarray}
F(x)
=\frac{2}{x^{2}} \left[  1-\frac{1}{\cosh ( \text{arcsinh} (x)/2)  }  \right]\quad.
\end{eqnarray}
Fig.~(\ref{sqttypeI}) demonstrates that the reduced quantity
\begin{equation}
\frac{S(q,T) }{S(0,T) } = \frac{F(q\xi_{||})  }{F(0)}=4F(q\xi_{||})
\label{redsqt}
\end{equation}
obtained in the continuum approximation
gives an excellent description of the discrete (TI) results
over many decades of the dimensionless
variable $q\xi$;  note that in the second equation
we have made use of the property $F(0)=1/4$.

\begin{figure}
\null\hskip 1truecm\psfig{figure=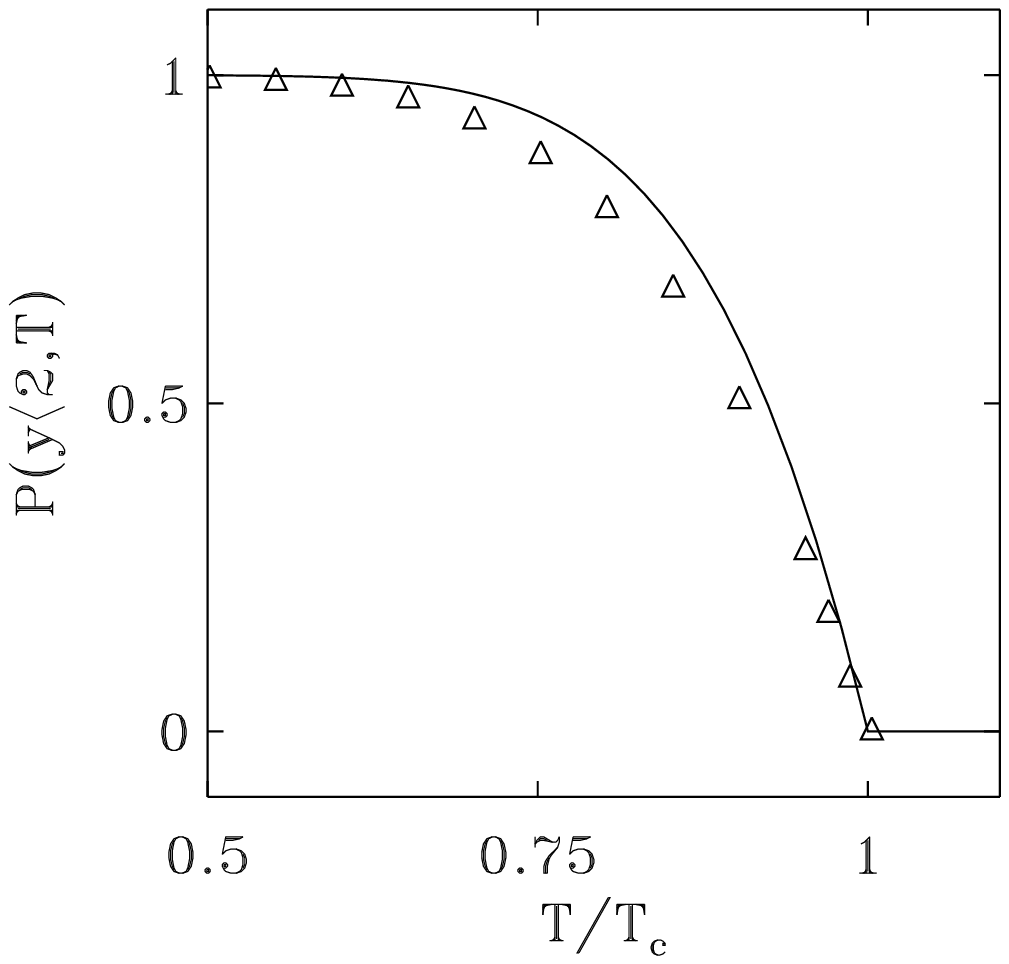,height=7truecm,width=7truecm}
\vskip 1truecm \caption{ The fraction of bound base pairs
$P(y<2\AA,T)$ as a function of the $T/T_c$. The triangles are TI
values, whereas the solid line corresponds to the theoretical
prediction~(\ref{defprobability}). } \label{probability}
\end{figure}

The zero stress isothermal susceptibility is given by
\begin{eqnarray}
\nonumber \chi &=& \lim_{h \to 0}
\left(\frac{\partial \sigma}{\partial h} \right)_{T} \\
\nonumber
& = &  \beta D a^2\- S(0,T)\\
\nonumber
&=&{1\over 4\delta^2}\left|t \right|^{-4}\\
&=&\left({1\over 2\delta_c}\ {T\over T_c}\right)^2 \left|t
\right|^{-4} \quad , \label{eqchi}
\end{eqnarray}
i.e.  the corresponding critical exponent is $\gamma=4$. Again
with $\delta_c=0.36$, Fig.~(\ref{sigmaandycarreI}) demonstrates
excellent agreement of (\ref{eqchi}) with TI values over many
decades.

At temperatures very near $T_{c}$ and finite $q$, i.e. for
$q\xi_{||}\gg1$, $F(x)\sim 2/x^{2}$ and hence
\begin{equation}
S(q,T_c)\propto {1\over (q\ell )^{2-\eta}}
\end{equation}
with $\eta=0$.

\subsection{The DNA order parameter}
The quantity which is directly accessible in
DNA denaturation in the fraction of bound base
pairs, which can be measured using UV absorbance.
The probability
$P(y<b,T)$ of finding a given base pair at an equilibrium distance
smaller than $b$ (equal to the fraction of bound pairs, with a proper
choice of $b$), is given by
\begin{eqnarray}
P_I(y<b,T) & =&  \int_{-\infty}^{b} dy \ |\phi_{0}(y)|^2 \\
&=& 1- \frac{\gamma(2\delta-1,2\delta e^{-ab}) }{ \Gamma(2\delta-1) }\\
& \approx&   (2\delta-1)\ Ei(2\delta e^{-ab})
\label{defprobability}
\end{eqnarray}
where $\gamma$ is the  incomplete gamma function, $Ei$ is the
exponential-integral function, and the last line is valid in the
vicinity of $T_{c}$. For the class of models effectively described
by the Morse potential, the fraction of bound pairs approaches
zero linearly as $T \to T_{c}$, for any value of $b$ (cf.
Fig.\ref{probability}). The slope, as expected, reflects the
particular choice of $b$.

\section{An alternative view of the transition: thermal stability of
the domain wall}
\subsection{Elementary dynamics}

The equation of motion corresponding to (\ref{eq:eHamiltonian}) is
\begin{equation}
m{\ddot y}_{n}=K (y_{n+1}  + y_{n-1} -2 y_{n}) - \frac{ \partial V}{\partial y_{n} }
\end{equation}
or, in the continuum limit,
\begin{equation}
{\ddot y} = c_{0}^{2 }\frac{ \partial ^{2 }y}{\partial x^{2 } }
-\frac{ 1}{m } \frac{ \partial V}{\partial y }
\label{EMcont}
\end{equation}
where $c_{0}=\omega _{0}\ell$. Relevant static configurations of
the infinite
chain with free ends, derivable from  (\ref{EMcont}) are:\\
(a) The uniform solution at the minimum of $V(y)$, $y(x) \equiv 0$; this
configuration corresponds to the absolute minimum of
the total potential energy (equal to zero) and is stable with respect to
small amplitude fluctuations; linearized solutions of (\ref{EMcont})
are optical phonons. \\
(b) For very large values of $y$, $V(y)$ is almost a constant;
therefore, expressions which verify $d^2 y /d x^2 = 0$, i.e. of
the form $  \lim_{M \to \infty} [ y(x) - M ] = C \cdot x $, where
$C>0$ is an arbitrary constant, are approximate solutions in the
sense that they correspond to local minima of the total potential
energy; linearization of
(\ref{EMcont}) around them leads to an acoustic phonon spectrum. \\
(c) an exact, unbounded, domain-wall like solution
\begin{equation}
y_{DW}^{\pm}(x) = \frac{ 1}{ a} \ln\left [ 1 + e^{\pm (x-x_{0})/d}  \right]    \quad.
\label{DWsol}
\end{equation}
where $d=\ell/ \sqrt{2R}$ and $x_{0}$ is an arbitrary constant.
\begin{figure}
\null\hskip 1truecm\psfig{figure=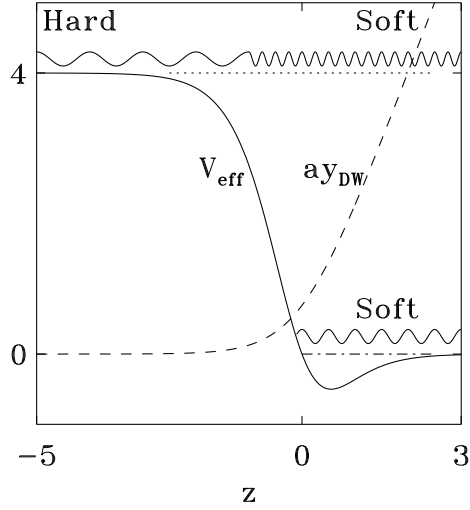,height=7truecm,width=7truecm}
\vskip 1truecm \caption{The dashed line represents the domain wall
solution (\ref{DWsol}) as a function of the dimensionless space
variable $z$. The full line represents the effective potential of
the Schr\"odinger-like equation (\ref{lin}). The potential
marginally fails to support  bound states. Scattering states with
eigenvalues less than the top of the potential (i.e. between
dashed and dashed-double-dotted lines) are confined to the
right-half of the line (low-frequency acoustic phonons); those
with higher eigenvalues consist of both transmitted and reflected
waves with different wavevectors and hence different dispersion
relations in the two halves of the line (corresponding to both
hard, i.e. optical, and soft, i.e. acoustic phonons, cf. text).}
\label{DWgraph}
\end{figure}

The solution (\ref{DWsol}) is plotted in Fig.~\ref{DWgraph} for
the case of the upper sign. It represents a  nonlinear field
configuration which "interpolates" from the stable minimum
(a) to a particular member of the metastable configurations (b),
with a slope $C =1/(ad)$, and hence equal contributions to the
elastic and the on-site potential energy densities ($D$ per site).
In other words it is a profile of a double chain where the two
strands stick together if $x<x_{0}$, but strand separation grows
linearly for all points $x \gg x_{0}$;  the energy of the
solution contains a term which is proportional to the number of
sites to the right of $x_{0}$. More exactly, if lattice sites are
numbered from $0$ to $N$,
\begin{equation}
E_{DW}^{+}=\left( N -\frac{ x_{0}}{ \ell}\right)2D + {\cal O}(N^{0 })  \quad.
\label{DWen}
\end{equation}
At zero temperature  the profile (\ref{DWsol}) is not stable. Since
\begin{equation}
E_{DW}^{+}(x_{0}\pm \ell) - E_{DW}^{+}(x_{0}) =\mp 2 \ell D
\end{equation}
the wall will spontaneously move to the right, "zipping" back the
unbound portion of the double chain. We will consider below how this
instability changes under the influence of temperature.
\subsection{Linearisation around the DW}
Consider small deviations with respect to (\ref{DWsol}), i.e.
\begin{equation}
y(x,t) = y_{DW}^{+}(x-x_{0}) +
\sum_{j} \alpha _{j} f_{j}(x-x_{0})e^{-i\omega _{j}t}
\end{equation}
where $|\alpha _{j}|\ll a$.  The linearized eigenfunctions $f_{j}$
satisfy the equation
\begin{equation}
-\frac{d^{2 }f_{j}}{ dz^{2 }} + 2 \biggl[1-\tanh z- {\rm sech}
^{2}z \biggr]f_{j} = \frac{ 2\omega _{j}^{2 }}{R\omega _{0}^{2 } }
f_{j}
 \quad,
\label{lin}
\end{equation}
where $z=(x-x_{0})/(2d)$. Eq.~(\ref{lin}) is Schr\"odinger-like
with an effective potential drawn in Fig.~\ref{DWgraph}. It can be
shown~\cite{MF} that (\ref{lin}) has no bound states; scattering
states are of two types:

(I) if $\omega _{q}^{2}<2R\omega _{0}^{2 }$
they are confined to the right half of the
$z$-axis; for $x\gg x_{0}$
\begin{equation}
f_{q} \sim \cos \left[k_{-}z+\delta _{I}\right]
\equiv  \cos \left[q(x-x_{0}) + \delta_{I} \right]
\end{equation}
where
\begin{equation}
\frac{ \delta _{I}}{2 }=\arg \Gamma \left(ik_{-}\right) -
\arg \Gamma\left(\frac{1}{2}\kappa _{+} -1 + \frac{i}{2}k_{-}\right)
-\arg \Gamma\left(\frac{1}{2}\kappa _{+} +2 + \frac{i}{2}k_{-}\right)
\label{phshI}
\end{equation}
$k_{-}^{2}\equiv 2\omega _{q}^{2 }/(R\omega _{0}^{2 })$,
$\kappa ^{2 } \equiv 4-k_{-}^{2 }$, $q\equiv k_{-}/(2d)$ and therefore
\begin{equation}
\frac{ \omega _{q}}{\omega_{0}  } = q\ell \quad \mbox{ if $q\ell <
\sqrt{2R}$} \label{softcontinuumphonon}
\end{equation}
i.e. the "phonons" with the low frequencies are soft, acoustic
phonons, as one expects them to be, since they correspond to atoms
oscillating on the flat portion of the Morse potential.

\begin{figure}
\null\hskip 1truecm\psfig{figure=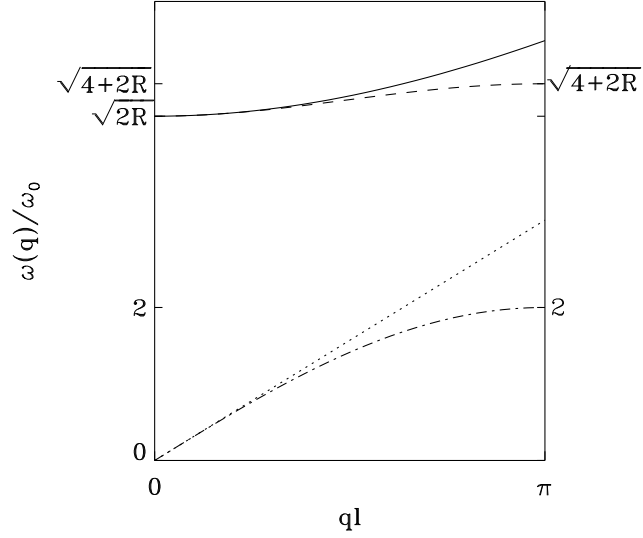,height=7truecm,width=7truecm}
\vskip  1truecm \caption{Continuum limit dispersion relations for
the soft (Eq.~(\ref{softcontinuumphonon}) as a dotted line) and
hard (Eq.~(\ref{hardcontinuumphonon}) as a solid line) phonons;
they extend to $ ql \to \infty $. The discrete dispersion
relations (\ref{discretephonon}) are also shown, with dashed line
(hard) and dash-dotted line (soft).   All curves have been drawn
for $R=10.1$; note the formation of a frequency gap between the
two branches of the discrete spectrum for $R>2$.}
\label{relationdispersion}
\end{figure}

(II) in the high frequency case, i.e. if
$\omega _{q}^{2 } > 2R\omega _{0}^{2 }$,
scattering states
extend over the whole line and have both transmitted and reflected
components.  Specifically,
\begin{equation}
f_{q}(z)   = \left\{ \begin{array}{lll}
  & e^{-ik_{-}z }  +  \rho^{1/2}e^{i(k_{-}z+\delta _{II}^{\rho}) }
  &  \mbox{if  $z\to \infty$}  \\
  &  \left( \frac{k_{-}}{k_{+}} \tau \right)^{1/2} e^{-i(k_{+}z+\delta _{II}^{\tau}) }
  & \mbox {if $z\to -\infty$}
\end{array}
\right.
\end{equation}
where
\begin{equation}
\rho =\left[  \frac{\sinh \frac{ \pi }{2 }\left( k_{-}- k_{+}\right) }
                       {\sinh \frac{ \pi }{2 }\left( k_{-}+k_{+}\right)} \right] ^{2}
\end{equation}
$\tau =1- \rho$, $k_{+}^{2 }=4-k_{-}^{2 }$,  $q_{+}\equiv k_{+}/(2d)$
and therefore

\begin{equation}
\frac{ \omega _{q}}{\omega_{0}  }
= \left\{ \begin{array}{lll}
     & q\ell &\mbox{if $q\ell>\sqrt{2R}$, $x\gg x_{0}$    } \\
    & \left[ 2R + \left(q_{+}\ell \right)^{2} \right]^{1/2}
& \mbox{any $q_{+}$, $x\ll x_{0}$      }
\end{array}
\right.\label{hardcontinuumphonon}
\end{equation}
i.e. the dispersion relation is different in the soft (unbound)
segment and the hard (bound) segment of the line (see
Fig.~\ref{relationdispersion}). Formally, we can combine the two
results for the dispersion into two branches, one optical and one
acoustic, for each value of q, without any restrictions. However,
it should be borne in mind that the acoustic phonons physically
reside in the soft (unbound) and the optical ones in the hard
(bound) segment of the line.

For completeness we list the phase shifts
\begin{eqnarray}
\delta_{II}^{\tau}& = & 2\arg \Gamma \left[  2+\frac{ i}{2 }
\left( k_{-}+ k_{+}\right)\right] -
\arg \Gamma \left(1+ik_{+}\right)
-  \arg \Gamma \left(1+ik_{-}\right) - \pi \\
\delta_{II}^{\rho}& = & -2\arg \Gamma \left[  2+\frac{ i}{2 }
\left( k_{-}+ k_{+}\right)\right] - 2\arg \Gamma \left[  2+\frac{
i}{2 } \left( k_{-} - k_{+}\right)\right] + 2  \arg \Gamma
\left(1+ik_{-}\right) \quad.
\end{eqnarray}

\subsection{Free energy of the thermally dressed DW}

At finite temperatures, the domain wall is accompanied
by a phonon cloud. The phonon cloud contributes
three terms to the free energy, which arise from
cases I (soft), II (soft), II (hard) presented in the previous
subsection. Because these phonons reside in different
- and in general unequal - portions of the chain,
their contributions to the free energy depend on the
position of the DW. Specifically,
\begin{eqnarray}
F_{soft} & = & k_{B}T \sum_{q}^{} \ln \beta \hbar \omega^{ac}_{q} \\
F_{hard} & = & k_{B}T \sum_{q}^{} \ln \beta \hbar \omega^{opt}_{q}
\label{f_hardsoft}
\end{eqnarray}
where contributions from soft modes have been combined.
The sum over q can be replaced by a density of states determined
by
\begin{eqnarray}
q\left(   N  - x_{0}\right) & = & n\pi \quad, n=0,1,2,... \mbox{(soft)}\\
q\left(   N + x_{0}\right) & = & n\pi \quad, n=0,1,2,... \mbox{(hard)}\quad.
\end{eqnarray}
This results in a phonon-cloud free energy
\begin{equation}
F_{phonon\> cloud} =  k_{B}T x_{0}\int _{0}^{\infty  }\frac{ dq}{\pi  }
\ln \frac{\omega^{opt}_{q} }{\omega^{ac}_{q} } + ...
\label{phcl}
\end{equation}
where the ellipsis denotes terms independent of $x_{0}$ and, since the
integrand is ultraviolet-convergent,   we have
extended the upper limit of the integration to infinity. This is
consistent with the continuum limit treated in this work; an
ultraviolet divergence present in the ellipsis can be corrected
by the introduction of a lattice cutoff but this is irrelevant for
the purposes of the present argument.
Introducing the correct dispersion relations for optical
and acoustic phonons, we can evaluate \cite{Grad} the integral in  (\ref{phcl}).
This results in a total free energy (DW plus phonon cloud)
\begin{equation}
F_{DW} = const + \left( k_{B}T\frac{ \sqrt{2R}}{2 } -2D  \right)\frac{ x_{0}}{\ell } \quad.
\label{FDW}
\end{equation}

\begin{figure}
\null\hskip 1truecm\psfig{figure=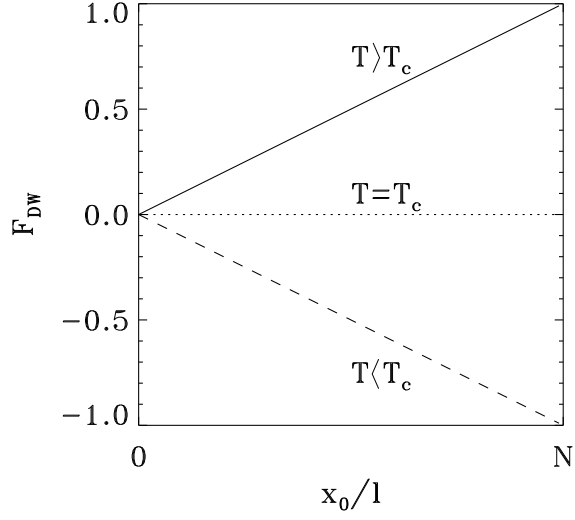,height=7truecm,width=7truecm}
\vskip 1truecm \caption{Free energy of the DW (in units of D) for
temperatures above ($T=3/2 T_{c}$, solid line), below ($T=1/2
T_{c}$, dashed line) and at $T_{ c}$ (dotted line), according to
Eq.~(\ref{FDW}); the irrelevant constant has been omitted. }
\label{freeenergyDW}
\end{figure}

Equation (\ref{FDW}) is the central result of this section.
It describes in very simple terms why and when the phase
transition occurs. At temperatures lower than
\begin{equation}
T_{c} = \frac{2\sqrt{2} D  } {k_{B}\sqrt{R}} = \frac{ 2\sqrt{2KD} }{ak_{B}}
\label{Tcsol}
\end{equation}
the prefactor of $x_{0}$ is negative; the DW's natural
tendency is  towards high positive values of $x_{0}$, i.e.
it "zips" the system back to the bound configuration.
Conversely, at temperatures higher than $T_{c}$,
thermal stability is achieved by a high negative value
of $x_{0}$, i.e. the DW "opens up", unbinds the system.
At the critical point, the DW is thermally (meta)stable at
any position; this corresponds to physical realizations
of the chain at criticality.  It should be noted that the
value of  $T_{c}$ predicted by the above DW argument
coincides with the exact result (\ref{Tcexact})
of the continuum limit. This supports the validity
of the alternative description of
the phase transition in terms of the thermal
stability of the DW. \par

\section{Discussion}

We have given a detailed account of the exact thermodynamics and
the scattering function $S(q,T)$ of a model used to describe
thermal DNA denaturation, as well as other one-dimensional
instabilities. The model deserves special attention for two
reasons: (a) it is probably the simplest exactly solvable lattice
model in one dimension which exhibits a true thermodynamic
transition and satisfies all scaling laws (cf. below), and (b) the
transition can be understood in terms of the thermal stability of
a soliton-like nonlinear configuration. A number of comments are
in order:
 \enumerate \roman{enumi}
\item None of the "prohibitions"
of phase transitions in one dimension apply to this model.
The theorem by Gursey~\cite{gursey}
and the van-Hove demonstration~\cite{vanHove},
extended by Ruelle~\cite{ruelle} apply to
systems with pair interactions only. The standard
Landau \cite{Landau} argument, which covers systems with on-site
potentials (e.g. $ \phi ^{4}$) rests on the finiteness of the
DW energy.

\item  The critical exponents calculated analytically
from the properties of the Schr\"odinger equation (\ref{eqmorse})
satisfy all applicable scaling laws, i.e $d\nu_{||}=2-\alpha =
\gamma + 2\beta$ and $\gamma = (2-\eta)\nu _{||}$.  Figs. 2-4
demonstrate very good  agreement between the thermodynamic
behavior calculated from numerical TI, and the one obtained
exactly from Eq.~(\ref{eqmorse}). It appears that "universality"
extends to prefactors and not just to exponents - provided that
corrections due to the absolute value of critical point are taken
into account. This in spite of the fact that the TI values  refer
to a system with a very high degree of discreteness ($R=10.1$).

\begin{figure}
\null\hskip 1truecm\psfig{figure=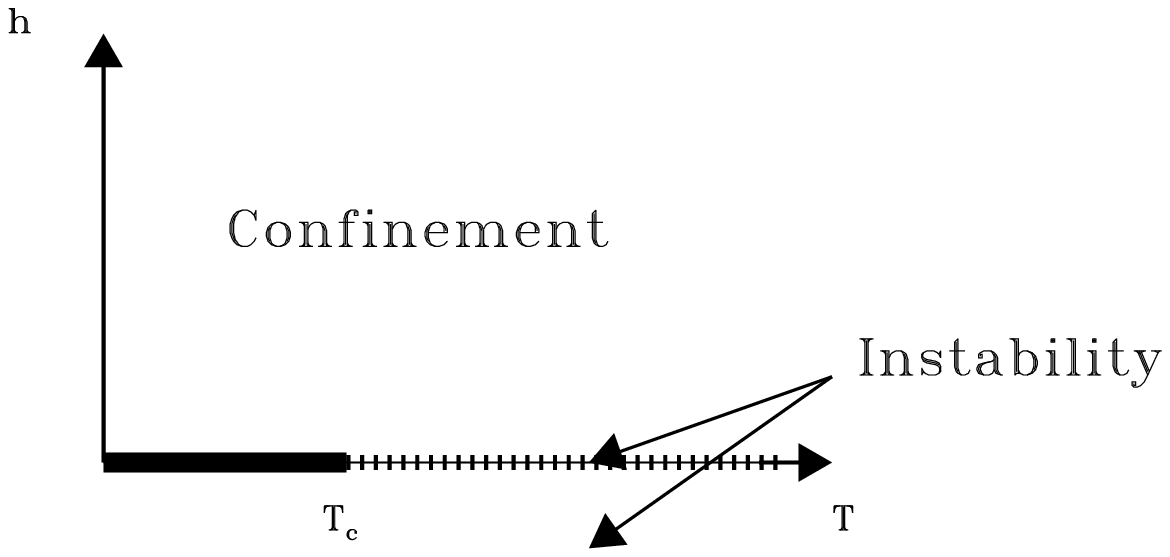,height=7truecm,width=7truecm}
\vskip 1truecm \caption{Phase diagram in the $(T,h)$ plane. At
negative $h$, the system is unstable at all temperatures.}
\label{phasediagram}
\end{figure}

\item It is instructive to view the phase
diagram in the $(h,T)$ plane (Fig. \ref{phasediagram}).
For positive values of
the external field $h$, there is confinement at all
temperatures. In the limit $h \to 0^{+}$, the system
approaches a confined ($T<T_{ c}$), or a
deconfined  ($T>T_{ c}$) state. The TI
(and the exact, Schr\"odinger)  solutions presented
here refer to the limit of the ($h=0^{+}, T$)
confined state as $T \to T_{ c}^{-}$
(thick line). Finally, at any $h<0$, the system is
unstable.

\item The scattering function (\ref{defsqtypeI}) shows typical
Ornstein-Zernike
behavior. Its form demands that at least some of the weight cannot
come from phonons. We will report details of critical dynamics in a separate
paper. Preliminary results suggest the occurrence of a
central peak. It will be interesting to relate critical dynamics
with large-scale fluctuations of nonlinear configurations.
\item The estimate~(\ref{Tcsol}) of the critical temperature can be extended
to  cover discrete systems. This is done by introducing the discrete
dispersion relations
\begin{equation}
\frac{\omega _{q}^{2 }}{\omega _{0}^{2 }}
=  \left\{   \begin{array}{lll}
      & 2 \left( 1- \cos(q\ell) \right) + 2R & \quad \mbox{(optical)} \\
      & 2 \left( 1- \cos(q\ell) \right)          & \quad \mbox{(acoustic)}
\end{array}
\right.
\label{discretephonon}
\end{equation}
in Eq.~(\ref{phcl}) which describes the free energy of the phonon
cloud accompanying the DW. Numerical experiments have demonstrated
\cite{PDT} that the  asymptotic slope of the DW is given
by the same function of $R$ for arbitrary levels of discreteness,
i.e. $\lim _{n \to \infty}
(y_{n}-y_{n-1})/\ell=1/(ad)=\sqrt{2R}/(a\ell)=
 \sqrt{2D/K} /\ell$.  It follows that each site which finds itself in the high
temperature phase contributes an elastic energy  equal to
$(1/2) K(y_n - y_{n-1})^{2} =D $
and an on-site energy equal to $D$, i.e. a total of
$2D$. This is a property of the DW, continuum or discrete.
Therefore, the only modification to (\ref{FDW}) and (\ref{Tcsol})
comes from the entropy of the phonon cloud. In detail,
using \cite{Grad}
\begin{equation}
I^{*}(R) \equiv \frac{ 1}{2\pi  } \int_{ 0}^{\pi} dx \ln [1-\cos x + R]=
\ln \left[ \frac{\sqrt{R}+\sqrt{R+2} }{2 }  \right]
\end{equation}
we obtain from (\ref{phcl})
\begin{equation}
F_{phonon\> cloud} = \frac{ k_{B}T x_{0}} {\ell} \left[ I^{*}(R)-I^{*}(0) \right]
\end{equation}
which results in the estimate
\begin{equation}
T_{c}=\frac{ 2D}{k_{B}\ln\left[ \sqrt{R/2} + \sqrt{1+R/2}  \right] } \quad.
\label{Tcdiscrete}
\end{equation}

\begin{figure}
\null\hskip 1truecm\psfig{figure=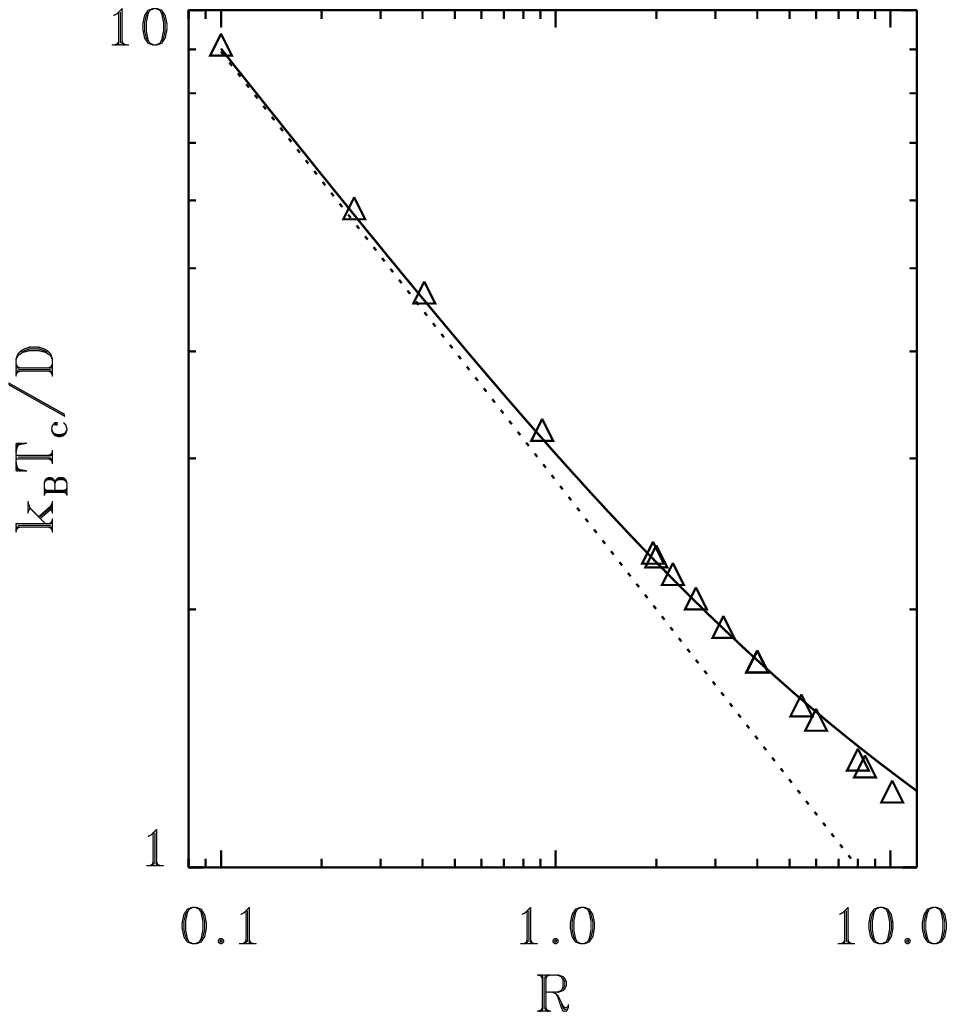,height=7truecm,width=7truecm}
\vskip 1truecm \caption{Dependence of the critical temperature on
the discreteness parameter $R$. The dotted curve shows the
continuum estimate,
 ~(\ref{Tcsol}). The solid curve
corresponds to ~(\ref{Tcdiscrete}), whereas the triangles denote
the numerical TI values; agreement is excellent for values of $R$
well into the discrete regime; nonetheless, systematic deviations
can be seen, starting  at $R>4$ (cf. text).} \label{evolutiontc}
\end{figure}

In the limit $R\ll 1$, Eq.~(\ref{Tcdiscrete}) reduces to
Eq.~(\ref{Tcsol}). It can be seen in Fig.~(\ref{evolutiontc})
that the {\em nonperturbative} theoretical estimate
(\ref{Tcdiscrete}) is in excellent agreement with TI numerical
results even for values of $R$ well into the discrete regime,
$R\gg1$. Systematic discrepancies appear at values of $R>4$ and
seem to grow as $R$ increases further. The origin of these
discrepancies is currently under investigation; preliminary
numerical results \cite{PDT} suggest a significant complexity in
the properties of the DW at values of $R>4$.
\endenumerate
Within the scope of the present work, we feel it is justified to
state that (a) in the continuum approximation, and (b) even at
moderate levels of discreteness, our understanding of a prototype
one-dimensional phase transition can be considerably enhanced by
making use of the concept of the thermal stability of a distinctly
nonlinear entity. In more plain terms: formation of the DW can be
thought of as "driving" the thermodynamic instability.

\acknowledgments We thank P. C. W. Holdsworth and  D. Mukamel for
helpful discussions.    This work has been partially supported
by EU contract No. HPRN-CT-1999-00163 (LOCNET network).

\end{document}